# Protonic Nickelate Device Networks for Spatiotemporal Neuromorphic Computing

Yue Zhou[1], Shaan Shah[1], Tamal Dey[2], Yucheng Zhou[1], Ashwani Kumar[1], Sashank Sriram[1], Siyou Guo[3], Siddharth Kumar[2], Ranjan Kumar Patel[2], Eva Y. Andrei[3], Ertugrul Cubukcu[4], Shriram Ramanathan[2], Duygu Kuzum[1*]

**Affiliations:**

[1]Department of Electrical and Computer Engineering, University of California, San Diego, California 92093, United States

[2]Department of Electrical and Computer Engineering, Rutgers University, Piscataway, New Jersey 08854, United States

[3]Department of Physics and Astronomy, Rutgers University, Piscataway, New Jersey 08854, United States

[4] Department of Chemical and Nano Engineering, University of California, San Diego, California 92093, United States

*Corresponding author. Email: dkuzum@ucsd.edu

**Abstract:** Computation in biological neural circuits arise from the interplay of nonlinear temporal responses and spatially distributed dynamic network interactions. Replicating this richness in hardware has remained challenging, as most neuromorphic devices emulate only isolated neuron- or synapse-like functions. In this work, we introduce an integrated neuromorphic computing platform in which both nonlinear spatiotemporal processing and programmable memory are realized within a single perovskite nickelate material system. By engineering symmetric and asymmetric hydrogenated $NdNiO_3$ junction devices on the same wafer, we combine ultrafast, proton-mediated transient dynamics with stable multilevel resistance states. Networks of symmetric $NdNiO_3$ junctions exhibit emergent spatial interactions mediated by proton redistribution, while each node simultaneously provides short-term temporal memory, enabling nanoseconds-scale operation with an energy cost of ~0.2 nJ per input. When interfaced with asymmetric output units serving as reconfigurable long-term weights, these networks allow both feature transformation and linear classification in the same material system. Leveraging these emergent interactions, the platform enables real-time pattern recognition and achieves high accuracy in spoken-digit classification and early seizure detection, outperforming temporal-only or uncoupled architectures. These results position protonic nickelates as a compact, energy-efficient, CMOS-compatible platform that integrates processing and memory for scalable intelligent hardware.

**Main Text:**

Neuromorphic computing aims to create intelligent systems that mimic the brain's way of processing information[1-5]. In biological neural networks, self-organization and interactions are essential for developing both structure and function, enabling adaptability, fault tolerance, and robustness[6,7]. Neural circuits distributed across different brain regions communicate through synaptic networks as well as spatiotemporally diverse brain waves to process and integrate neural information needed for complex cognitive capabilities[8-10]. Current hardware approaches to neuromorphic computing, typically focus on individual components such as discrete neuron or synapse devices rather than the network as a whole. This contrasts with the complex topology of biological computation systems, where the interaction between structure and function is critical for emergent computational capabilities. Here, "emergent" refers to system-level collective behaviors arising from spatiotemporal interactions among coupled nodes, and not present in the response to inputs to any individual node. As a result, such approaches may fall short in capturing the emergent collective dynamics of biological neural networks, where complex computation naturally arises from self-organizing spatiotemporal interactions among interconnected neurons.

Inspired by the recurrent connectivity of biological neural networks[11,12], recent efforts[13-17] have harnessed the intrinsic nonlinear temporal dynamics of memory devices to implement reservoir-like computing architectures[18,19]. However, these approaches predominantly focus on isolated devices exhibiting temporal dynamics alone, generally characterized by millisecond-scale switching and comparatively higher energy consumption, while fall short of capturing the network-level interactions. In-material neuromorphic platforms have shown spatial coupling but their full functionality often depends on external processors and off-chip components, limiting system-level integration[20-24]. To date, no single material system has combined nonlinear temporal dynamics, emergent spatiotemporal behavior and nonvolatile switching characteristics required for such integration. Such homogeneous integration minimizes compatibility concerns between processing and output layers, thereby offering the potential to simplify the fabrication process for large-scale integration.

In this work, we implement a homogenously-integrated neuromorphic computing system based on a perovskite nickelate, NdNiO$_3$ on the same wafer. Perovskite nickelates demonstrate a range of electronic phenomena, including metal-insulator transitions, electronic switching[25], magnetic ordering[26], and have been explored as building blocks for artificial neurons, synapses, and memory capacitors[25,27]. However, energy-efficient transient switching in the nanoseconds regime and the integration of spatiotemporal processing with nonvolatile outputs within a single material system have not been explored. Here, by engineering symmetric and asymmetric hydrogenated NdNiO$_3$ (H-NNO) junctions on the same wafer, we realize volatile time-dependent dynamics for nonlinear spatiotemporal processing together with non-volatile programmable switching for output layer in a unified platform. Spatiotemporal interactions between nodes emerge from proton-mediated dynamics, facilitating the self-organized formation of recurrent connections. Our approach leverages ultrafast protonic dynamics (~500 ns) and emergent connectivity in H-NNO films, creating an energy efficient (0.2 nJ per input) and brain-inspired computational system capable of real-time data processing without fine-tuning of individual nodes.

**Spatiotemporal computing in hydrogenated perovskite nickelates**

The computational framework consists of two distinct layers: a dynamic spatiotemporal processing layer and a static, programmable output layer (**Fig. 1a**). The spatiotemporal processing layer utilizes the nonlinear temporal dynamics in each node and recurrent connectivity between each node to nonlinearly map input signals $u$(t) into a rich internal state representation $x$(t). These interactions and the individual dynamics help capture temporal dependencies in the input and generate a rich set of dynamic features, which allows a simple linear model to classify or predict the output effectively. Our processing layer exhibits dynamic emergent connectivity (W$_{dynamic}$) spontaneously evolving with the input signals while the output layer employs static, pre-trained weights ($W_{static}$) to linearly combine internal states $x(t)$ and generate stable outputs $y(t)$. **Fig. 1b** illustrates the corresponding hardware implementation integrated on a H-NNO film. In the spatiotemporal layer, each Pd electrode corresponds to a physical node. When voltage spike trains are applied to Pd electrodes, each node exhibits temporal fading memory and interacts with other nodes through emergent spatial coupling mediated by the inputs and the proton redistribution in the H-NNO film. The resulting output current from each Pd electrode is converted to a voltage signal and used as the input to the linear output layer. In the output layer, Pd-Au pairs with programmable resistance implement the output weights ($W_{out}$). The voltage signals from the Pd-Pd processing array are applied to the Pd electrodes, and the currents collected at the Au electrodes form the outputs ($y_1, y_2, ..., y_m$) for the computational tasks.

Both the spatiotemporal processing layer and the linear output layer are fabricated onto a single NNO film grown on LaAlO$_3$ (LAO) wafer (**Fig. 1c, Fig. S1**). Fabrication process flow is outlined in the **Methods** section and **Fig. S2**. Optical and scanning electron microscopy images of the Pd-Pd spatiotemporal processing arrays and Pd-Au linear output arrays are shown in **Fig. 1d-g**. Upon annealing in hydrogen gas, Pd acts as a catalytic electrode and dopes hydrogen into the nickelate lattice, forming H-NNO[25,27-31]. In contrast, Au acts as an inert electrode establishing a hydrogen gradient in the channel. Cu-Kα X-ray diffraction (XRD) analysis (**Fig. S3**) showing broadening and shifting of NNO peak (002) confirm hydrogenation consistent with the literature[32]. The symmetric Pd-Pd devices (**Fig. 1h**) consists of two hydrogen clouds moving in opposite directions, results in a transient current response that accumulates over repeated voltage pulses, exhibiting short-term memory. In contrast, the asymmetric Pd-Au devices (**Fig. 1i**) have a single hydrogen cloud proximal to the Pd electrode. Application of voltage pulses cause migration of protons surrounding the Pd electrode leading to a long-term resistance change.

Compared to established in materia reservoir computing (RC), our H-NNO platform features dual-timescale protonic dynamics: (1) non-volatile proton incorporation provides long-term memory states that can be stably programmed, and (2) volatile capacitive proton motion introduces short-term dynamics with tunable decay. In addition, our H-NNO networks differ both in interaction mechanism and controllability. While conventional in materia RC relies on fixed, nearest-neighbor coupling, our system exhibits substrate-mediated global interactions in which each Pd electrode reshapes the voltage potential landscape and influences all other nodes. At the same time, every node is electrically accessible allowing programming and direct current readout, providing flexible control of network connectivity, which is absent in conventional in-materia RC systems, delivering a richer and more flexible spatiotemporal computational framework.

**Non-volatile switching and transient dynamics in nickelate devices**

The resistance modulation in hydrogenated nickelate devices is largely due to proton migration under applied electric fields[27]. During hydrogen annealing, Pd catalyzes the interstitial doping of hydrogen into the NNO lattice near the electrode. The hydrogen atoms donate electrons to the Ni $d$ orbitals, which shifts the nickel's oxidation state and creates a phase transition, drastically reducing electrical conductivity[25,33-35]. Electrochemical impedance spectroscopy (EIS) and transport characteristics (**Fig. S4, Supplementary Text 1**) confirm the metal-to-insulator transition in H-NNO. Similar proton-induced resistance modulation is also observed in NNO films fabricated on $SiO_2$/Si substrates, indicating compatibility with silicon platforms (**Fig. S5**). **Fig. 2a** shows the switching characteristics of Pd-Au devices, featuring only one hydrogen cloud. Positive bias on the Pd electrode exhibits a RESET (resistance increase), while negative bias produces a SET (resistance decrease). Kelvin probe force microscopy (KPFM) imaging has shown that resistance modulation is a result of proton redistribution near the Pd electrode under voltage bias[32]. The asymmetric I-V response with a larger SET window than RESET, is consistent with Schottky barrier formation at the H-NNO/NNO interface, as discussed in **Supplementary Text 2** and **Figs. S6-S8**. **Fig. 2b** shows 16 distinct non-volatile resistance levels of the Pd-Au device. The highly linear I–V relationship is suitable for implementing matrix-vector multiplication in linear arrays. The change in device resistance as a function of the number of applied pulses (**Fig. 2c**) shows that as the number of positive pulses increases, the resistance of the Pd-Au device gradually increases due to the migration of the hydrogen cloud towards NNO film, whereas the negative pulses move the hydrogen cloud back to the Pd electrodes. Consistent pulse update is observed across devices (**Fig. S9**). The stability of 16 resistance states over 1000 seconds (**Fig. 2d**) confirms the non-volatility of the programmed states. A larger number of states (i.e., 36 states, **Fig. S10**) can be achieved by spanning a wider resistance range (~40–350 kΩ). Extended retention of representative states up to 10,000 s (**Fig. S11**) further verifies their long-term stability.

The symmetric Pd-Pd device under DC sweep reveals a lower absolute current and smaller switching window (**Fig. 2e**). The lower absolute current could be explained as the presence of two hydrogen clouds forming back-to-back Schottky barriers at the H-NNO/NNO interfaces[29] (**Fig. S12**). The smaller switching window originates from competing resistance modulation: under both positive and negative voltage sweeps, one hydrogen cloud expands while the other shrinks. Because hydrogen shrinkage (resistance decrease) is stronger than expansion (resistance increase), a slight final resistance decrease is observed for both sweep polarities. Hydrogen migration in both asymmetric Pd-Au and symmetric Pd-Pd junctions is further confirmed by KPFM measurements (**Fig. S13**, **Supplementary Text 3**). When a voltage is applied between the two Pd electrodes, the local potential of the NNO film is determined by resistive voltage division between the two hydrogen clouds beneath the Pd electrodes. Under nanoseconds voltage pulse trains, this local potential of the NNO film gradually increases due to capacitive charging of the LAO substrate (**Fig. S14**). This dynamic voltage potential affects the local voltage experienced by each hydrogen cloud, further modifying their distribution. As shown in **Fig. 2f**, the transient current response accumulates over repeated pulses and relaxes after pulse removal due to the discharging, exhibiting short-term memory behavior. **Fig. 2g** shows the current decay following a voltage pulse (5 V, 500 ns) which can be fitted by an exponential decay with a time constant of 5 μs. This explanation of observed transient behavior is supported by circuit simulations (**Supplementary Text 4, Fig. S15**), using experimentally measured H-NNO resistance and parasitic capacitance values (**Fig. S16**).

To further investigate the temporal memory effect, multiple pulses with the same configuration but different time intervals varying from 500 ns to 3.5 μs are applied (**Fig. 2h**). Stronger accumulation effect is observed under pulse trains with shorter time intervals. Such frequency-dependent behavior could be described in two steps: fast integration process during pulse application and slower exponential relaxation after pulse removal. The corresponding model equations and fitting procedures are provided in **Supplementary Text 5**. The dependence of the accumulation strength and decay time constant on pulse amplitude is summarized in **Fig. S17**. Statistical analyses of Pd-Au devices and Pd-Pd devices are performed to evaluate device uniformity and stable switching behavior (Supplementary **Figs. S18-S25** and **Supplementary Text 6**). Geometric scaling characteristics, including the effects of electrode spacing and pad size, are presented in **Supplementary Fig. S26-28** and **Supplementary Text 7**, demonstrating scalable and tunable spatiotemporal dynamics in Pd-Pd devices.

**Spatial interactions in H-NNO arrays**

Symmetric Pd-Pd devices arranged in a two-dimensional array enable spatial interactions through proton redistribution in the NNO film. When an electric pulse is applied to one node, proton redistribution modifies the local hydrogen cloud resistance, which in turn changes the global voltage potential of the NNO film and impacts nearby nodes. **Fig. 3** experimentally investigates representative features of input-mediated emergent spatial interactions by monitoring the output current of a reference device ($D_{C,R}$) under controlled neighboring activity. These measurements provide the basis for modelling spatial coupling for large-scale spatiotemporal H-NNO network simulation. In all experiments, $D_{C,R}$ is driven by a fixed voltage pulse train (5 V amplitude, 500 ns width, 500 ns interval), while neighboring devices are independently configured to receive the same pulse train or remain grounded. The current measured at the grounded terminal of $D_{C,R}$ increases and saturates over time, with its magnitude dependent on the activity of surrounding nodes. **Fig. 3a-3b** investigates the in-plane coupling effect on $D_{C,R}$ due to neighbouring devices in four different directions ($D_{C,R-1}$, $D_{C+1,R}$, $D_{C,R+1}$ and $D_{C-1,R}$). The results consistently demonstrate that the presence of a positive pulse train in a neighbouring device leads to a higher output current in the reference device compared to when the neighbouring device is grounded. The impact of neighboring devices shows weak dependence on physical distance distances (**Fig. 3c-d**) Notably, the device $D_{C,R+4}$, located furthest from the reference device, produces the largest output current impact when a positive spike train is applied. This behavior is attributed to the the large conductivity contrast (~ 3 orders of magnitude) between the metallic NNO substrate and the hydrogenated H-NNO regions (**Fig. S4d**), which causes the voltage drop across the entire NNO film to remain negligible in comparison to the voltage across the H-NNO clouds. As a result, the coupling strength is determined by the resistance ratio of the hydrogen clouds between the neighboring and reference devices, together with the applied pulse configuration, rather than by the physical distance, yet still reflects substrate-mediated spatial interactions in the nickelate.

We further examine a 2 × 3 array to study the collective influence of multiple neighboring devices (**Fig. 3e**). **Fig. 3f** shows that configurations with two positive pulse trains in neighboring devices yield the highest output current, followed by one positive pulse train. Conversely, the all-ground configuration produces the lowest output. This demonstrates that the output of the reference device is amplified depending on the level of spiking activity in its surrounding nodes. This effect is nonlinear due to variations in hydrogen clouds across the array, nonlinear temporal switching characteristics of individual devices and emergent coupling through the substrate.

To validate these findings, COMSOL simulation is performed (**Methods**). The simulated current distribution within a single Pd electrode and the corresponding effective resistance are presented and discussed in detail in **Fig. S29** and **Supplementary Text 8**. Furthermore, **Fig. 3g** illustrates the voltage potential distribution in a 2 × 3 device array under different configurations. Depending on the resistance of individual hydrogen clouds and the applied voltages, the potential landscape across the NNO substrate is reshaped, which modulates the local electric field acting on each hydrogen cloud, thereby coupling their dynamics. Specifically, when two positive spike trains are applied to neighboring devices, the voltage difference experienced by the reference device becomes the largest, resulting in the highest output current, consistent with our measurement results. Building on the COMSOL simulation and experimental observations, the substrate-mediated spatial coupling can be further understood through a quantitative formulation that describes the pairwise coupling resistance between electrodes. The detailed derivation is provided in **Supplementary Text 9**.

**Hardware implementation of spatiotemporal computing**

For a complete hardware implementation of a pattern recognition task, we connected the spatiotemporal processing and linear output layer arrays (**Fig. 4**). Symmetric Pd-Pd spatiotemporal processing array performs nonlinear feature transformation, while asymmetric Pd-Au array performs linear weighted sum operation for the output layer. The Au pad integrates the current from the surrounding Pd pads (**Fig. S30** and **Supplementary Text 10)**. The four input patterns U, C, S, and D are represented as 5×5 binary matrices (**Fig. 4a**) and converted into voltage spike trains. An example for the S pattern is shown in **Fig. 4b**, where black elements correspond to voltage pulses with 500 ns pulse width and 5V amplitude and white elements represent ground. These spike trains are subsequently applied to the Pd pads in the spatiotemporal processing array (**Fig. 4c).** Pad 6 is grounded and serves as the reference electrode, while the other pads receive the encoded inputs. Signal transfer between the spatiotemporal Pd-Pd layer and the linear Pd-Au output layer is implemented off-chip. The output currents from Pad 5 and Pad 6 are first readout and externally converted into voltages ($V_{out1}$ and $V_{out2}$), which are then applied to the linear output array as inputs. Voltages with opposite polarity are applied to realize negative weights in the hardware, enabling accurate linear mapping for pattern recognition.

The real-time monitored current responses from Pad 5 ($I_1$) and Pad 6 ($I_2$) are shown for each pattern (**Fig. 4d**). The U and D patterns share similar temporal evolution but exhibit different current amplitudes. For the C pattern, a pulse at the last time step prevents full current decaying, resulting in a higher residual current. For the S pattern, consistent pulses are applied at every time step, producing continuous current accumulation. The distinct voltage distributions among patterns (**Fig. 4e**) demonstrate that the spatiotemporal processing layer encodes pattern-specific features. The trained weight distribution of the Pd-Au output array is shown in **Fig. 4f**. Applying these weights to the spatiotemporal outputs produces distinct current responses across output columns for each pattern (**Fig. 4g**), demonstrating successful classification. These results illustrate that the integration of spatiotemporal processing with linear weighted sum operations achieves effective pattern recognition. We further examine the feasibility of on-chip integration by directly interconnecting the spatiotemporal and linear output layers. As shown in **Fig. S31**, simultaneous probing allows voltage pulses to be applied to the spatiotemporal layer while monitoring the current response in the linear output layer. The measured outputs (**Fig. S32**) follow the induced current trends under 500 ns pulsing. It confirms that the signal can be reliably transferred across layers even with the additional load and parasitic capacitances from a direct connection.

We further design and perform large-scale neural network simulations for speech recognition and neural signal classification tasks (**Fig. 5**). **Fig. 5a** shows the spoken digit recognition task pipeline using the AudioMNIST dataset[36]. Audio signals are encoded as voltage spike trains, processed by the Pd-Pd spatiotemporal processing layer, and subsequently classified using a Pd-Au linear output layer. **Fig. 5b** compares the evolution of hydrogen cloud thickness changes in the Pd-Pd array under temporal-only and spatiotemporal processing. In the temporal-only case, hydrogen cloud thickness evolves according to nonlinear temporal decay at individual nodes. In contrast, under spatiotemporal processing, the thickness of each hydrogen cloud is additionally modulated by spikes from neighboring channels through protonic coupling, facilitating information propagation beyond individual nodes. Detailed dataset descriptions, preprocessing steps, and implementation parameters are provided in the **Supplementary Information text 11**.

Using the generated current features as inputs to the linear output layer array for linear regression, we classify the spoken digits and evaluated the accuracy. **Fig. 5c** compares classification accuracy for three cases: without processing layer, temporal only processing, and spatiotemporal processing. The accuracy improves as the number of sampling times increases in all three cases, as more information is provided to the networks. Spatiotemporal processing achieves the highest accuracy (95.3%) outperforming both temporal-only and regression with no processing (uncoupled) baselines. We investigate the accuracy increase by studying pairwise distance across input samples after spatiotemporal processing (**Fig. S33**). Spatiotemporal processing exhibits the highest pairwise distance compared to the temporal-only and no-processing layer cases validating enhanced separation between digit classes compare with temporal-only and no-processing cases (**Supplementary Text 12**). We assess the energy efficiency of our approach relative to other computing platforms, (**Supplementary Table 1** and **Supplementary Text 13)**. The reported energy values account only for the intrinsic device operation within the H-NNO layer, excluding peripheral readout or front-end circuitry. Compared to the RC hardware based on ionic motion (millisecond timescales), our protonic system provides substantial energy saving while operating at 500 ns timescales.

To further demonstrate the broader applicability and advantages of our protonic neuromorphic hardware, we evaluate neural signal processing of electroencephalogram (EEG) signals for seizure detection. We use the CHB-MIT Scalp EEG Database (version 1.0.0) available from PhysioNet[37-39] which contains EEG recordings sampled at 256 Hz from 23 electrodes placed according to the International 10-20 system. A representative EEG recording from a single patient, containing both seizure and normal events, is shown in **Fig. 5d**. EEG signals are binarized into spike trains using a fixed threshold and processed by the spatiotemporal layer, followed by a linear output layer, to classify the events as either seizure or normal. A representative normalized seizure event from one patient is shown in **Fig. 5e**, with corresponding non-seizure signal clip presented in **Fig. S34**. Threshold optimization is shown in **Fig. S35**. Across all threshold settings, the spatiotemporal processing layer consistently outperformes both the temporal-only and no processing baselines. To evaluate the performance for early seizure detection which is critical for timely intervention, we further restrict the input to the first 1, 2 and 3 seconds of EEG signals. Even with limited input, spatiotemporal processing achieves the highest seizure detection accuracy, 85%, in contrast to 58% for no processing and 67% for temporal only processing (**Fig. 5f**).

**Conclusion**

We demonstrate a neuromorphic computing platform in perovskite nickelates by integrating symmetric Pd-Pd junctions as a spatiotemporal processing layer with asymmetric Pd-Au junctions as programmable output layers on a single wafer. Spatial coupling between hydrogen clouds enables nonlinear feature transformation and encoding, while short-term memory dynamics capture temporal correlations in input signals. Leveraging these dynamics, the platform achieves emergent spatiotemporal interactions and efficient real-time classification of temporal signals.

We implement large-scale spoken digit recognition and early seizure detection tasks to establish the advantages of protonic neuromorphic computing with perovskite nickelates for energy efficient real-time temporal data processing. Beyond these demonstrations, the emergent spatiotemporal dynamics in H-NNO networks may also be relevant for broader applications, including audio signal processing, video feature encoding, and biosignal monitoring and analysis in biomedical systems.

Furthermore, heterogeneous integration of nickelates with other material systems may also open promising directions for spatiotemporal neuromorphic computing. For instance, strongly correlated oxides exhibiting metal-insulator transitions and thermal coupling could potentially serve as spatiotemporal processing layers[21], while ferroelectric materials with sub-nanosecond switching dynamics and ultra-low energy consumption may provide efficient output layers[40,41]. Perovskite nickelates thus offer a versatile platform for next-generation neuromorphic systems, bridging the gap between biological computing principles and hardware implementations.

**Acknowledgments**

The network fabrication was supported by "Quantum Materials for Energy Efficient Neuromorphic Computing," an Energy Frontier Research Center funded by the US Department of Energy (DOE), Office of Science, Basic Energy Sciences, under Award DE-SC0019273 (DK and SR). DK acknowledges the Office of Naval Research (N00014-24-1-2127) for simulations, computations, and analysis. SG and EYA acknowledge DOE-FG02-99ER45742 and Gordon and Betty Moore Foundation GBMF9453 for AFM and KPFM experiments. The material characterization was supported by AFOSR Grant FA9550-22-1-0344 (SR).

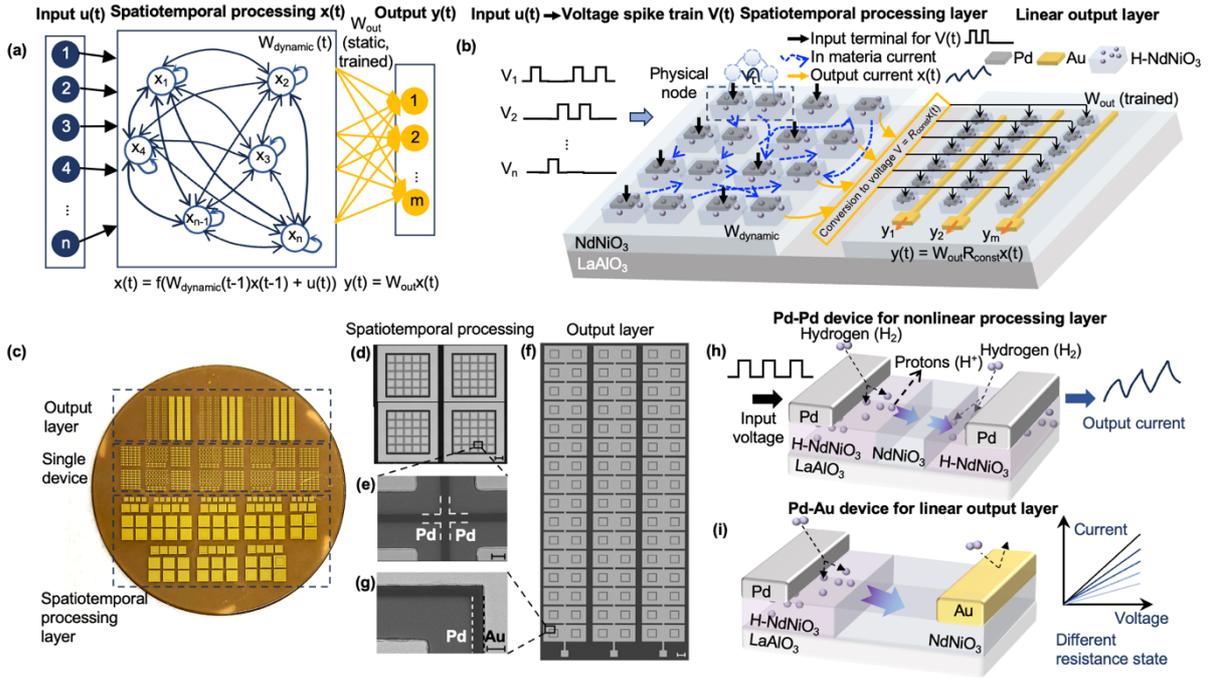

**Fig. 1: Perovskite hardware for efficient data processing and computation on the same wafer. (a)** Schematic of the computational framework, consisting of two distinct layers: a dynamic spatiotemporal processing layer and a static output layer. The equations at the bottom describe the evolution of the state $x(t)$ and the computation of the output $y(t)$, demonstrating the dynamic nature of the spatiotemporal processing layer's response and the linear mapping in the output. **(b)** Schematic of the corresponding hardware implementation integrated on a single $NdNiO_3/LaAlO_3$ substrate. The time-dependent input signals $u(t)$ is proportionally converted into voltage spike trains $V(t)$ are applied directly to catalytic Pd electrodes on a H-NNO film. The resulting output current x(t) from each electrode is monitored as the output of the spatiotemporal processing layer. **(c)** Photograph of the fabricated chips with single device, spatiotemporal processing layer and linear output layer units. **(d)** Microscope image of spatiotemporal processing arrays, the scale bar is 100 μm. **(e)** Zoomed-in scanning electron microscopy (SEM) image of spatiotemporal processing units, the scale bar is 5 μm. The distance between each Pd electrode is 3.5 μm. **(f)** Microscope image of output layer arrays, the scale bar is 250 μm. **(g)** Zoomed-in SEM image of a single device unit for output layer, the scale bar is 5 μm. **(h)** Schematic of the Pd-Pd nickelate devices, where one hydrogen cloud expands while the other shrinks, exhibiting time-dependent hydrogen redistribution and charge relaxation. **(i)** A schematic of the Pd-Au nickelate devices with hydrogen cloud migration from the Pd electrode enabling non-volatile resistance change.

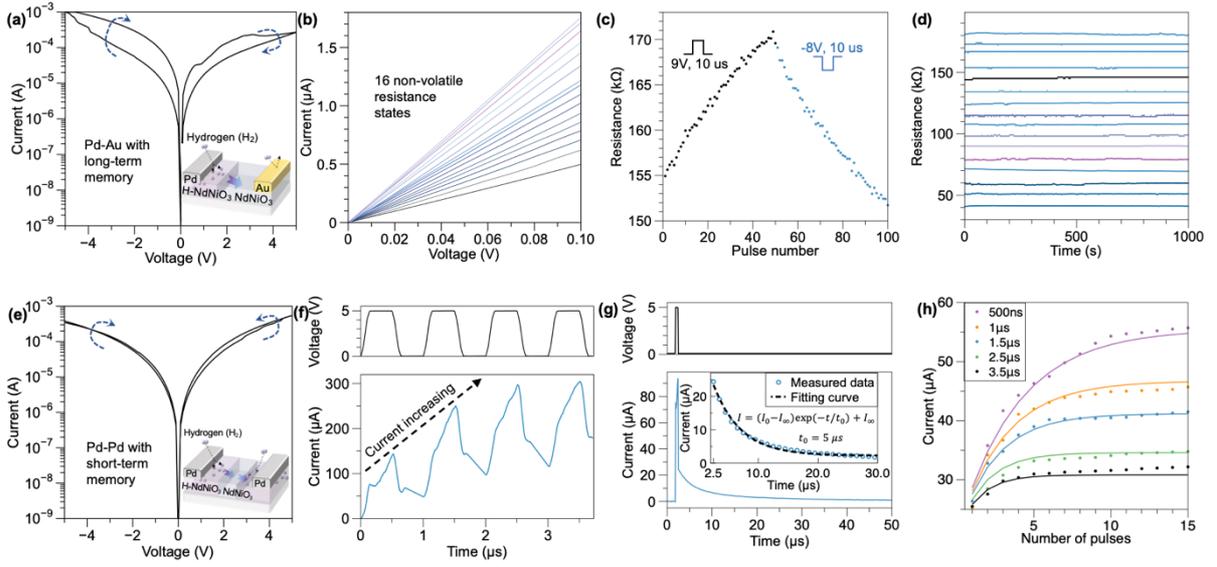

**Fig. 2: Device characteristics of non-volatile Pd-Au and dynamic transient Pd-Pd devices. (a)** I–V characteristics of Pd-Au devices under -5V to 5V sweep, showing a large hysteresis window. **(b)** I–V curves under a small read voltage sweep (0 to 0.1V) showing 16 distinct resistance states. **(c)** Non-volatile and programmable device resistance change under a series of positive (+9V) and negative (-8V) voltage pulses (10 μs in width), the device resistance is read under 0.1V. **(d)** 16 stable resistance states over 1000 seconds at 0.1V read voltage, indicating multi-level stability within a single device. **(e)** Pd-Pd device I–V hysteresis curves obtained from a dual voltage sweep ranging from -5V to 5V. **(f)** Real-time output current change of the device under periodic voltage signal input. **(g)** Output current as a function of time under single voltage pulse stimulation (5 V, 500 ns and constant 0.1 V voltage bias). **(h)** Gradual increase in device current following a train of voltage pulses (5V and 500 ns) with varying time intervals. The current is sampled immediately after the removal of each pulse using a 0.1 V bias. Higher pulse frequencies resulted in a more pronounced current increase. The dots represent experimental data and solid lines represent the fitting curves based on equation 3.

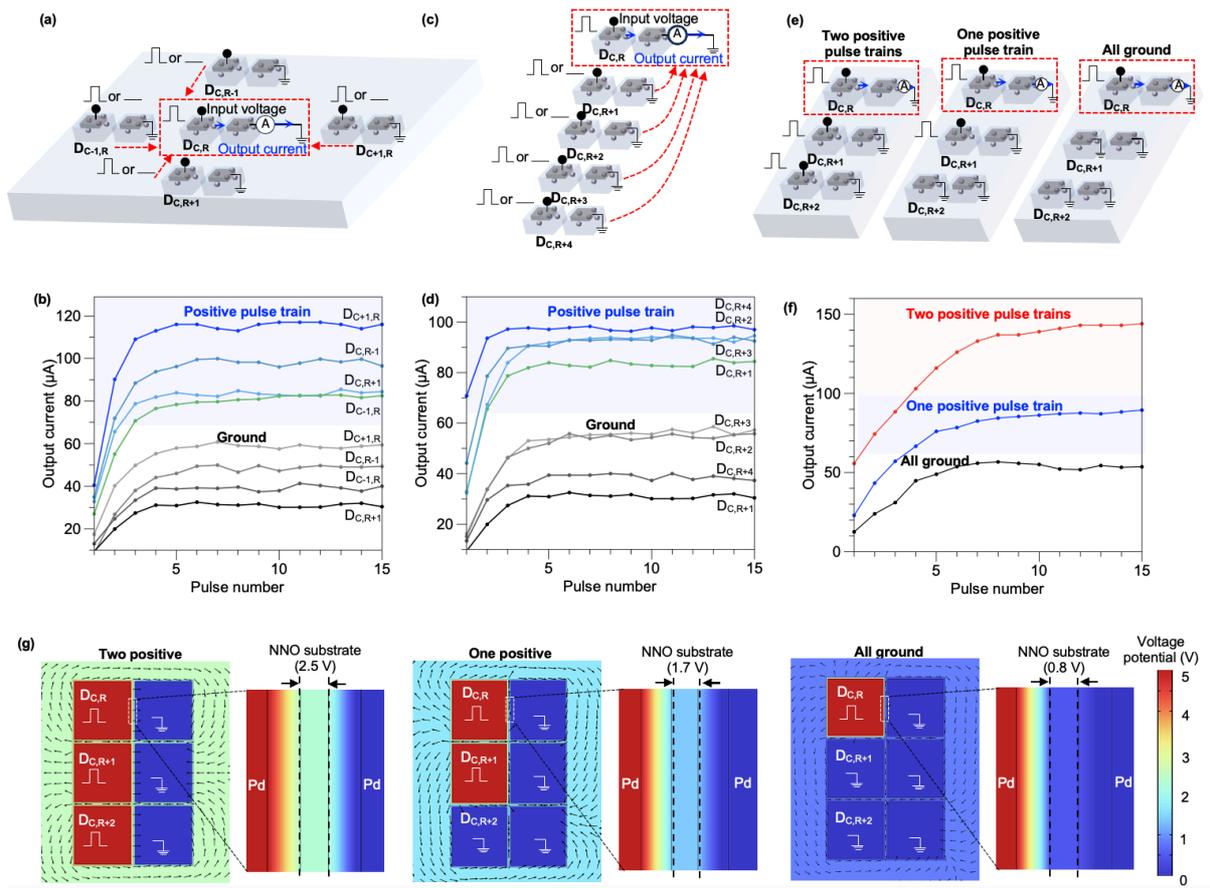

**Fig. 3: Spatial interactions between Pd-Pd nickelate devices. (a)** Schematic illustration of the coupling effect on the reference device $D_{C,R}$ from neighboring devices in four directions. A voltage pulse train (5 V, 500 ns pulse duration, and 500 ns time interval) or a ground signal is applied to the neighboring devices in each of the four directions. Meanwhile, a constant voltage pulse train (5 V, 500 ns pulse duration, and 500 ns time interval) is applied to the reference device, and the output current is measured from the reference device to evaluate the influence of directional coupling. **(b)** Experimentally measured output current of the reference device $D_{C,R}$ when a voltage pulse train or ground signal is applied to neighboring devices in four directions. **(c)** Schematic illustration of the coupling effect on the reference device $D_{C,R}$ from neighboring devices at varying distances. **(d)** Experimentally measured output current of the reference device $D_{C,R}$ when a voltage pulse train or ground signal is applied to neighboring devices at varying distances. **(e)** Diagram of spatial interactions between three Pd-Pd devices, illustrating three different stimulation configurations: two positive pulse trains, one positive pulse train, and all grounded. The diagram shows how these varying stimulation conditions influence the reference device $D_{C,R}$, demonstrating the combined effect of multiple

neighboring devices on the behavior of the reference device. (**f**) Experimentally measured output current of the reference device $D_{C,R}$ in three configurations. The configuration with all neighboring devices receiving a pulse train produces the largest output current. (**g**) COMSOL simulations of the voltage potential distribution under different neighbor configurations: two neighboring devices driven by positive pulses, one neighboring device driven by a positive pulse, and both neighboring devices grounded. The resulting potential distribution changes with the input condition of each device, confirming substrate-modulated spatial interactions.

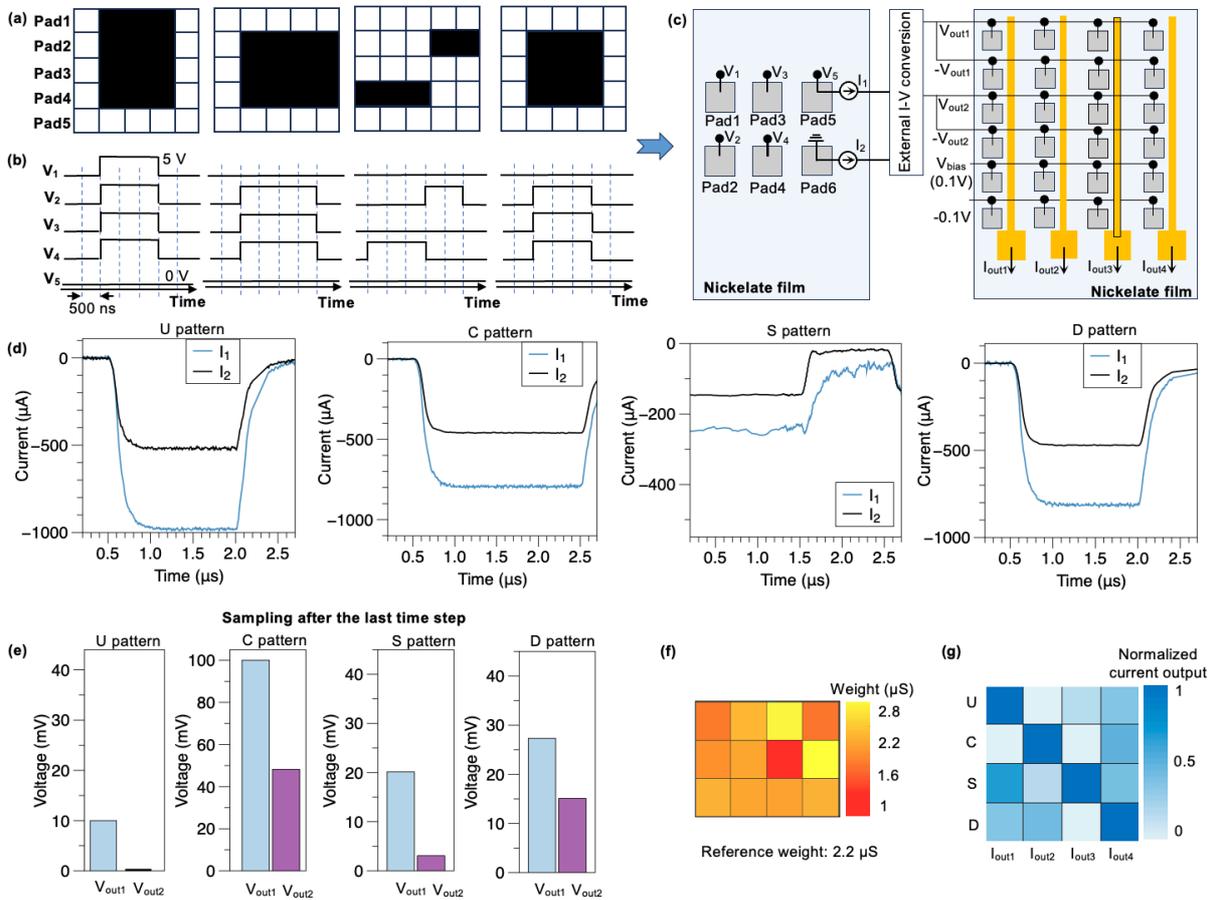

**Fig. 4: Experimental results of pattern classification using a fully perovskite nickelate neural network structure in hardware.** (**a**) Patterns (U, C, S and D) used for this experiment. (**b**) An example of the converted voltage spike trains corresponding to the S pattern. (**c**) The schematic of spatiotemporal processing layer and output layer array connection for a fully NNO-based network. (**d**) Real-time monitored current response from pad 5 and 6 when U, C, S, and D patterns. (**e**) The voltage applied on the linear output layer proportionally converted by the current output from pad 5 and pad 6. (**f**)

Experimental synaptic weights distribution of Pd-Pd spatiotemporal processing array. (**g**) Experimentally measured currents in output layer array with four columns.

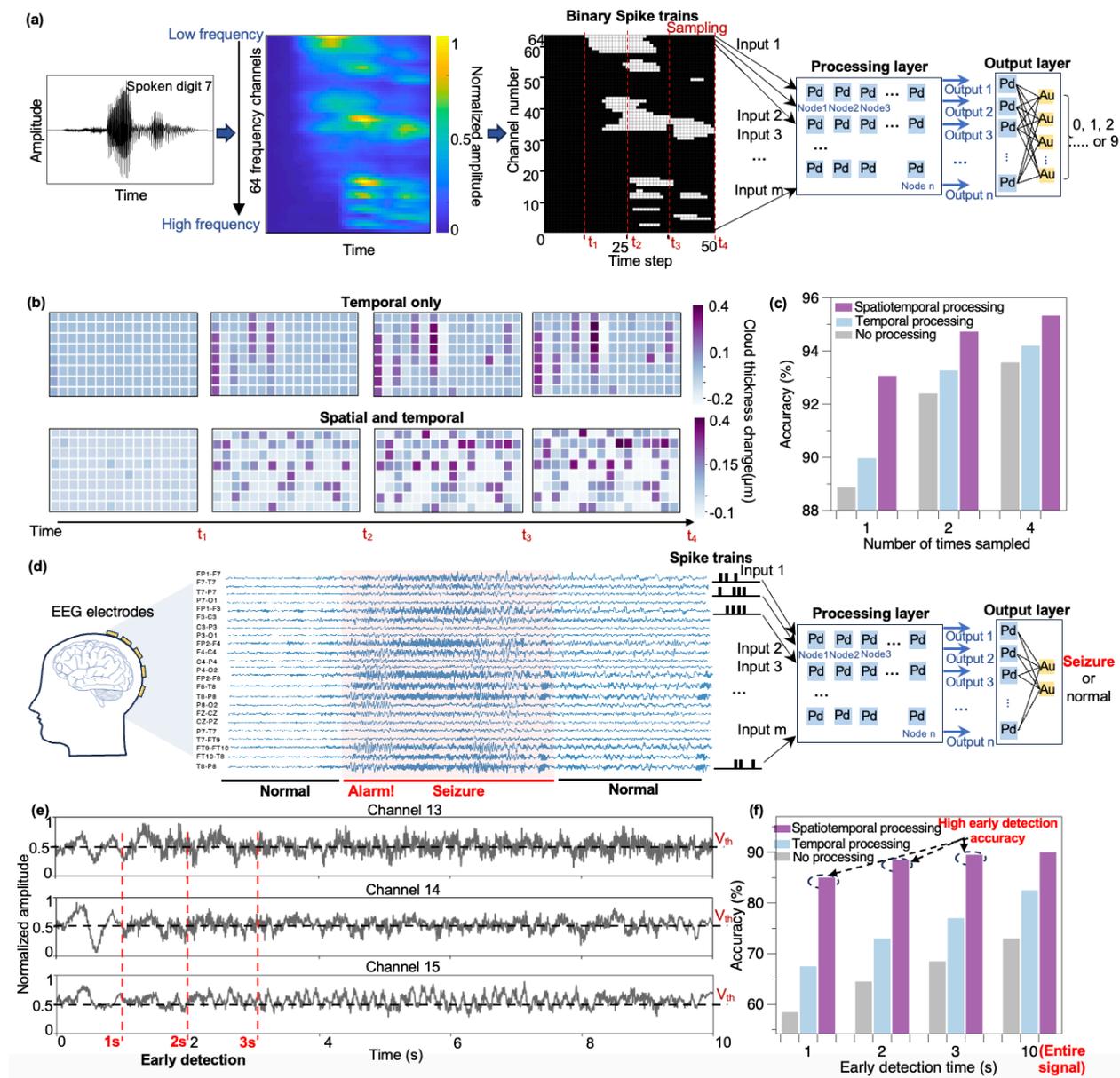

**Fig. 5: Perovskite nickelate networks for spoken digit recognition and seizure detection. (a)** Processing pipeline for spoken digit recognition task. The Lyon's Ear model, inspired by the human cochlea, is used to convert the audio signals into analogue signals with 64 frequency channels, which were converted into binary voltage spike trains via thresholding to be fed into the spatiotemporal processing layer. At each time step, a 5V voltage pulse with 500 ns width was applied whenever there was a spike. Spike trains processed by the preprocessing layer were applied to a linear output layer to classify the inputs into the correct digits. (**b**) Evolution of the hydrogen cloud thickness from 128 Pd electrodes. The results are compared for cases where every two electrodes are isolated with only temporal characteristics (temporal only) versus when all 128 electrodes exhibit spatial interactions (spatial and temporal). (**c**) Accuracy of speech recognition is plotted against number of times (1,2,4) the spike train is sampled for no pre-processing, temporal only and spatiotemporal cases. The results show that the spatiotemporal processing by protonic nickelate networks increases the digit recognition accuracy. (**d**) Processing pipeline for early seizure detection. EEG signals recorded from 23 electrodes placed on the scalp according to the International 10-20 system. The displayed signals include segments of both normal brain activity and seizure episodes. A threshold is applied to convert the continuous signal into spike trains and applied into the preprocessing layer classified by a linear output layer as either seizure or normal. (**e**) A representative seizure event lasting 10 s shown for three channels. Time points for early detection are marked as 1s, 2s and 3s. (**f**) Seizure detection accuracy results show that after the spatiotemporal processing by protonic nickelate networks offers a substation advantage for early detection of seizures with high accuracy.

**Methods**

**NdNiO$_3$ array fabrication**

NdNiO$_3$ thin film (50 nm) was deposited on a 2-inch LaAlO$_3$ (LAO) (100) single-crystal wafer (1 mm thickness) at room temperature using radio-frequency (RF) sputtering. The deposition process utilized co-sputtering of Ni and Nd targets with 75W DC and 125W RF power, respectively. Deposition was performed at 5 mTorr pressure maintained by Argon/Oxygen gas mixture with flow rates of 40 sccm and 10 sccm, respectively. After deposition, the wafer underwent post-annealing at 550 °C for 24 hours in 25 sccm O$_2$ gas flow to enhance crystallinity. A bilayer lift-off process (LOR5B and AZ1512) was employed to define Pd electrode patterns on the NNO/LAO substrate. Pd electrodes (50 nm) were subsequently deposited via sputtering. Ti/Au (10 nm/100 nm) electrodes were then patterned and deposited using the same bilayer lift-off process. During the Au deposition step, a smaller-area Au pad was also deposited on top of each Pd electrode to improve electrical contact during device measurements. After Au and Pd metal deposition, the samples were placed in a tube furnace and annealed at 115°C for 20 minutes in a Hydrogen/Argon (5%/95%) atmosphere at a flow rate of 35 sccm. This process led to hydrogenation of the $NdNiO_3$ (H-NNO) thin films beneath the Pd electrodes, forming the proton-modulated nickelate junctions used in the device array.

**Kelvin Probe Force Microscopy (KPFM)**

We have used an AFM instrument (NT-MDT Solver Next AFM) for the AFM and KPFM experiments. The Instrument consists of an HA_NC tip made of monocrystalline silicon. The curvature of the tip was less than 10 nm and it has a spring constant of 2.8 N/m. AC modulation and frequency values were 0.5V and 131 KHz. During a two-step scan process, topography of the samples recorded using AFM in tapping mode. In the next step, contact potential difference ($V_{CPD}$) between the tip (which was retracted by $\Delta z = 10$ nm) and the sample were recorded and mapped spatially.

**Electrochemical impedance spectroscopy (EIS)**

A Gamry Reference 3000 potentiostat was used to perform EIS studies on the devices. Since our HNNO devices are solid state devices with two electrodes, we have connected one of them to the working electrode of the potentiostat, and the other electrode to the reference and counter electrode of the potentiostat. The measurements were done in the frequency range of 10 Hz to 1 MHz. The DC voltage applied was of 1.5 V amplitude and a perturbation AC signal of 100mV was applied.

**Electrical measurement of $NdNiO_3$ array**

The electrical characterization of the NdNiO$_3$ array was performed in a FormFactor Summit probe station using a Keithley 4200-SCS semiconductor analyzer. A 4225-PMU module was employed to generate electric pulses ranging from 500 ns to 10 μs. For real-time, high-precision current measurements during voltage pulse application, a 4225-RPM remote amplifier was utilized. To monitor the relaxation process following a voltage pulse, a constant 0.1 V voltage bias was applied and the current was measured in real-time at 100 ns intervals. All electric pulses were applied to the Pd electrode while the Au electrode was kept ground. The device resistance was extracted by fitting the current-voltage curve within the linear low voltage regime (-0.1 to +0.1V). All measurements were conducted in air at room temperature.

**Equivalent circuit modelling of Pd-Au and Pd-Pd devices**

The circuit simulation based on Pd-Au and Pd-Pd devices was performed using Cadence. To emulate the hydrogen cloud movement beneath the Pd electrode, a Verilog-A compact model was developed, treating hydrogen migration as a resistive switching memory device with its resistance state modulated by applied voltage (positive RESET, negative SET). A voltage-driven resistance evolution mechanism was incorporated to capture the nonlinear I-V behavior during large voltage sweeps, incorporating a scaling factor to adjust the resistance dynamics. For Pd-Au devices with only one hydrogen cloud per device, the model employed a single resistive switching memory element to reflect unidirectional hydrogen migration. In contrast, Pd-Pd devices were modeled using two elements with opposite polarities connected in series to represent simultaneous hydrogen cloud expansion and shrinkage. The model, integrated with capacitive effects, was incorporated into a SPICE circuit simulation framework, enabling equivalent circuit analysis of both Pd-Au and Pd-Pd devices based on experimental pulse measurement data.

**COMSOL simulation for spatial interactions in 2×3 Pd-Pd arrays**

The spatial distribution of electric potential in the 2×3 Pd-Pd array was simulated using the COMSOL AC/DC solver module. The entire array was simplified as a 2D surface, with Pd electrodes modeled as conductive metal regions. The resistivity values for NNO and H-NNO were set to $2.5\times10^{-6}$ Ω·m and 8.85 Ω·m, respectively. The simulated geometry was based on experimental dimensions, where each Pd pad was a square of 120 μm in length, with a 10 μm interval between adjacent Pd pads. The hydrogen cloud beneath each Pd electrode was assumed to be uniformly distributed and extended 3.5 μm beyond each side of the electrode, resulting in a total hydrogen cloud length of 17 μm. By applying different voltage configurations to each Pd electrode, the potential distribution across the nickelate film surface was computed using the two-dimensional model.

**Large scale Pd-Pd spatiotemporal processing layer modeling**

For the large-scale neural network simulation of spoken digit recognition tasks, a 128-node Pd-Pd spatiotemporal processing array was implemented in Python. Each Pd node was initialized with a random hydrogen cloud thickness ($x$) between 2 and 2.5 μm, and its conductance ($G$) was calculated based on $x$ using the equation provided in **Supplementary Text 5**. The spatiotemporal state evolution was updated every time step (500 ns) through three key processes: voltage potential evolution, hydrogen cloud evolution and current readout. The voltage potential across the NNO film was influenced by capacitive charging and discharging effects, which depend on the number of Pd nodes receiving a voltage spike at a given time step. This process introduced both nonlinear temporal characteristics and spatial interactions between nodes. Simultaneously, the hydrogen cloud thickness at each node was dynamically updated based on the local electric field which is determined by the voltage difference between the Pd node and the NNO film, as well as the current cloud thickness. The combined effects of capacitive behavior and hydrogen migration result in the final current accumulation and relaxation behaviors which define the short-term memory and spatial interaction properties of the spatiotemporal processing layer. In the end, the system's spatiotemporal response was captured by measuring the output current $I_t$ under a small read voltage 0.1V applied to half of the nodes. The resulting spatiotemporal processing layer output of each spoken digit signal had a size of (1, 128 × $N_{sample}$), where $N_{sample}$ represents the number of times each signal is sampled.

**Large-scale Pd-Au Output layer training**

A single-layer linear regression model is used as the output layer for spoken digit classification, implemented in Python. The trained weight values are directly linear mapped to the conductance states of the Pd-Au output layer array for hardware realization. The total spatiotemporal processing layer output **X** has a size of ($N_{train}$, 128 × $N_{sample}$), $N_{train}$ represents the total number of training samples and $N_{sample}$ represents the number of times each signal is sampled. The target matrix **Y** has size of ($N_{train}$, 10), where each row is a one-hot encoded representation of the corresponding digit label.

To evaluate the spatiotemporal processing layer's performance, we use 5-fold cross-validation. The dataset is split into training (**X**$_{train}$, **Y**$_{train}$) and testing (**X**$_{test}$, **Y**$_{test}$) sets using an 80-20 split. This process is repeated five times, ensuring that each subset serves as the test set once while the remaining four are used for training.[32]

For linear regression, the relationship between input and output is modelled using weights **W** of size (128×$N_{sample}$, 10) and a bias vector $b$ of size (1, 10). The predicted class probabilities $\hat{Y}$ with size ($N_{train}$, 10) are computed as:

$$\hat{Y} = XW + b \qquad (1)$$

The final class label $\hat{y}$ with size ($N_{train}$, 1) is determined using the argmax operation:

$$\hat{y} = \arg\max_{i \in 1,2,3,4\ldots 10} \hat{Y}_i \qquad (2)$$

The optimal weights **W** and bias **b** are obtained by minimizing the Mean Squared Error (MSE) loss, defined as:

$$L = \frac{1}{N_{train}} * \sum_{i=1}^{N_{train}} \|Y_i - \hat{Y}_i\|^2 \qquad (3)$$

where $N_{train}$ represents the number of training samples.

**Seizure Detection**

CHB-MIT Scalp EEG Database (version 1.0.0) available from PhysioNet[37-39] including EEG recordings sampled at 256 Hz from 23 electrodes placed according to the International 10-20 system was used. Our training dataset included EEG data from five patients, each contributing 20 seizure and 20 non-seizure 10-second EEG segments (a total of 200 signal segments). For clear comparison, both normalized seizure and normal signal clips from channels 13 to 15 are presented in **Fig. S17**. The whole network setup remained consistent with that used in the spoken digit recognition task, except for a reduced pad array size (from 128 to 46) to accommodate the smaller number of EEG input channels (23 versus 64 in the spoken digit dataset). The classification accuracy among different threshold setup is shown in **Fig. S18**, where a 10-second EEG clip was sampled at 10 evenly spaced time points (i.e., sampling interval of 1 second per point). As observed, there exists an optimal threshold setting that maximizes classification accuracy, which is expected: if the threshold is set too low, both seizure and non-seizure signals may generate excessive activations, introducing noise and reducing accuracy. Conversely, if the threshold is too high, useful information that differentiates the two classes is filtered out, also degrading performance.